# ICT and Health System Performance in Africa: A Multi-Method Approach

**Lucas Mimbi**
School of Computing
University of South Africa
Science Campus, Johannesburg
Email: mimbilg@unisa.ac.za

**Felix Bankole**
School of Computing
University of South Africa
Science Campus, Johannesburg
Email: bankofo@unisa.ac.za

## Abstract

For the past two decades, the discussion regarding the effect of ICT on health systems is becoming apparent. However, past studies have mainly focused on ICT impact on specific social-economic phenomena. Little empirical research on ICT and health systems exists. Many African countries have invested in ICT and there is a need to examine if such investments have impacted on health system of these countries. Using a multi-method approach, data for 27 African countries were analysed. We employed Data Envelopment Analysis, Cluster Analysis and Partial Least Squares to examine the impact. The findings indicate that the 27 countries can be grouped into three clusters based on their relative efficiency scores of ICT and health systems. More compelling, the findings indicate that countries that performed efficiently in ICT inputs also did so in their health systems. Further, findings indicate that ICT significantly improves life expectancy at birth and reduces infant mortality rate. African countries must significantly invest in ICT to improve their health systems so as to achieve socio-economic development. The current study has theoretical, methodological and policy implications.

**Keywords**

ICT and Health Systems, Data Envelopment Analysis, Regression Analysis, Cluster Analysis, Africa

## 1   Introduction

In recent years, there has been much discussion on the impact of information communication technologies (ICT) on health systems in both developed and developing countries (Bloom and Standing 2008; Lucas 2008). However, little research that focuses on the performance of ICT on the health systems particularly in Africa exists (Deidda et al. 2014; Al-Lagilli et al. 2011). Health systems are fundamental in ensuring improved citizens' welfare and of nations as well (WHO 2014). Underdeveloped health systems not only hamper individual's social and economic development, but also they may cause a detrimental effect on national economic prospects (WHO 2014). The recent Ebola outbreak in Africa has caused economic downturn and almost brought to halt economic activities in the affected areas. Many of the affected countries could hardly contain the outbreak due to inadequate health care services. The Ebola outbreak thus calls for improved health systems if African countries could register the required economic, social and political development.

Studies have indicated that investment in ICT is related to health (Ngwenyama, Andoh-Baidoo, Bollou and Morawczynski 2006). However, empirical evidence on the impact of ICT on health is fragmented and often focuses on case-studies of specific ICT investment (Bankole et al. 2011). The high rate of ICT adoption in Africa has revolutionised the health sector. Health sectors in Africa have been experiencing a tremendous change as a result of transition from industrial age medicine to information age health care services due to ICT adoption. The purpose of this paper is to stimulate awareness about the impact of ICT on health systems in Africa. Therefore, our research main questions focus on whether or not the continuous investment in ICT can impact the health systems performance.

Although several studies have focused on evaluating the impact of ICT on health systems at both economic and organizational level in developed and transition economies (Jeremic et al. 2012; Osei-Bryson and Ko 2004), the issue of evaluating the efficiency of health systems in Africa has not been elaborated. Health is one of the dimensions of human development that is considered to be a fundamental contributor to the welfare of a nation (Bankole et al. 2011). It is therefore imperative to evaluate the impact of ICT adoption on the performance of country's health system.

We therefore propose to analyse the data of ICT and health in 27 African countries using multi-method approach: - Data Envelopment Analysis (DEA), Cluster Analysis and Partial Least Square Regression Analysis with Structural Equation Model. The objectives are to determine the efficiency, performance and the impact of ICT infrastructure on health systems in Africa.

The rest of the paper is organised as follows: In section 2 we present the conceptual background of the key concepts applied in this study. Section 3 discusses the theoretical framework guiding the study, followed by methodology in section 4. In section 5 we present the results followed by contribution of the study in section 6. Lastly, we conclude our study in section 7.

## 2    Conceptual Background

### 2.1    Information Communication Technologies (ICT) Outlook

ICT has penetrated in our community and at least transformed the state of affairs of human life. Nowadays people cannot escape the influence of ICT and many countries have adopted ICT (Gatautis 2015). Given the potential of ICT in transforming national economies and reducing poverty, in 2000 it was agreed that the UN Millennium Development Goals (MDGs) be implemented by 2015. As we come to the end of implementing MDGs, many countries have made progress in respect of ICTs such as mobile-cellular subscriptions, Internet use, fixed and mobile broadband services, home ICT etc. The ICT implementation in Africa looks good especially in mobile broadband subscriptions though still low and more efforts are required. It is estimated that by end of 2015 there will be a 17.4 percent penetration rate of mobile broadband from the mere 4 percent in 2011 (ITU 2011; ITU 2015). Currently, one in five people use the Internet in Africa, making a 20 percent Internet penetration. However, ICT development in Africa is behind that of Europe and America (see Table 1).

| Indicator | Europe | America | Africa |
|---|---|---|---|
| Percentage of households with Internet access | 82.1 | 60 | 10.7 |
| Percentage of individuals using the Internet | 77.6 | 66 | 20.7 |
| Mobile broadband subscriptions per 100 inhabitants | 78.2 | 77.6 | 17.4 |

*Table 1. Selected ICT indicators*

### 2.2    Health Systems in Africa

Health is an important facet of development and potentially may determine how individuals can actively get involved in productive activities and contribute to the welfare of the country (Al-Lagilli, Jeremic, Seke, Jeremic and Radojicic 2011). The World Health Organisation (WHO) has emphasised a need to improve health care services particularly in Africa (WHO 2014). WHO defines health as a state of complete physical, mental and social well-being and not merely the absence of disease or infirmity (WHO 1946). This means that without good health it is impossible to actively participate in activities that promote social and economic development at both micro and macro levels. The WHO Africa region report of 2014 clearly indicates that Africa is largely behind in economic development due partly to health related challenges such as HIV/AIDS, tuberculosis and malaria (WHO 2014). These challenges are further exacerbated by the shortage of health workers and health services delivery facilities. Consequently, people in Africa are engulfed in poverty. It is estimated that about 45 per cent of the 330 million people in Africa live on less than one US dollar a day (WHO 2014). Reports suggest that Africa lags behind other regions in most of the health indicators (Deaton and Tortora 2015) (see Table 2). For instance, life expectancy at birth for Africa in 2009 was 54 years on average, 14 years behind the World average of 68 years. This means that most African live shorter than their counterparts elsewhere in the World. Similarly, the physician-patient ratio is alarming. In 2010, there were 2.3 physicians for every 1000 people. This suggests that some people in Africa may spend their entire lifetime without receiving professional medical attention. Mortality rate also is alarming, 107 children for every 1000 live births died before their fifth birthdays in 2011 (WHO 2014).

| Selected Health indicators | | | | |
|---|---|---|---|---|
| Indicator | World | Africa | Europe | year |
| Life expectancy at birth | 68 | 54 | 75 | 2009 |
| Physicians per 1000 people | 14.0 | 2.3 | 33.3 | 2010 |
| Under 5 mortality per 1000 live births | 51 | 107 | 13 | 2011 |

*Table 2. Selected Health Indicators*

Given the poor health situation in Africa, there is a need for improving health systems to prevent both communicable and non-communicable diseases and create health nations with health people who can actively participate in economic activities. The 2014 WHO report for Africa states that

> "African countries will not develop economically and socially without substantial improvements in the health of their people. The health-care interventions — treatments, diagnostic and preventive methods — that are needed in this Region are known. The challenge for African countries and their partners is to deliver these to the people who need them, and the best way to do this is to establish well-functioning health systems" (WHO 2014, p. xiii).

It is believed that most deaths related to diseases in African region are due to diseases that are preventable, treatable or both (Deaton and Tortora 2015). These deaths could be avoided if basic health care services and efficient health systems to deliver them were widely available (WHO 2014). How will Africa establish well-functioning health systems? The next section presents the discussion on the potential of ICT in ensuring well-functioning health systems in Africa.

## 2.3 ICT and Health Systems

The world has seen a proliferation of ICT adoption in provision of health care services (Qureshi, Kundi, Qureshi, Akhtar and Hussain 2015). The assumption is that ICT would significantly improve health systems and ensure even people in underserved areas are able to receive health services (Kwankam 2004; Mars and Scott 2015). The next section presents ICT implications for health care services.

## 2.4 eHealth and mHealth

The term eHealth could be described as the use of ICT in hospitals (Kimaro 2006). eHealth is defined as the use of ICT in provision of health care services (Kwankam 2004). This means ICT can be used in various health care functions such as clinical, educational, research and administrative regardless of geographical settings (Kwankam 2004; Mars and Scott 2015). mHealth extends the efficiency and accuracy of the already available health systems through the use of electronic devices such as PDAs and mobile telephone networks to improve functions (such as reporting procedure) of the health systems (Lucas 2008). ICT can be used to transform the health paradigm by shifting the provider-patient configuration (Lucas 2008). This typically can involve extending health care services to underserved areas by use of electronic or telecommunication means (Telemedicine) such as video chat, or health telephone hotlines. This arrangement provides patients with seamless access to doctors, improved diagnosis and treatment around the clock (Durrani et al. 2012; Lucas 2008; Shaqrah 2010). eHealth and mHealth can transform the health systems by incorporating electronic means to deliver information and provide health related training. For example, mobile phones can be used to disseminate information regarding vaccination campaign and the Internet can be used to provide distance learning, share information etc. (Lucas 2008; Qureshi et al. 2014).

mHealth occur mostly through mobile provision of health care services. It occurs through wireless telemedicine that involve the use of mobile telecommunications and multimedia technologies and their integration with mobile health care delivery systems (Istepanian and Lacal 2003). mHealth continues to mature and has been used to address health care challenges such as access, quality, affordability, matching of resources and behavioral norms through mobile technologies (Qiang, Yamamichi, Hausman, Altman and Unit 2012). mHealth is a network involving people and products; and the mechanisms that connect them using digital technologies. For instance, mobile health worker visits households in rural Botswana and educates family members about HIV/AIDS using mobile phones (WHO 2014). This shows that the potential of ICT in ensuring efficient health systems need to be investigated.

## 3 Theoretical Framework

In the present study we adopted the Cobb Douglas Production Function (CDPF) to guide the investigation of the impact of ICT on health systems. Over the years, CDPF has become a useful tool in growth theories (Stijepic 2015). CDPF has been widely used in studies involving ICT impact in various environments (e.g., Lee, Gholami and Tong 2005; Samoilenko 2008; Bankole, Osei-Bryson and Brown, 2013). Given its analytical simplicity and consistent in production function of growth accounting, we adopted CDPF in the present study. The impact of ICT Infrastructure on Health System performance in Africa can be represented by production function as follows:

i) The Impact of Mobile Communication Services on Health System = *f*[Mobile Cellular Subscribers (MCS) + Life Expectancy at Birth (LEB) + Infant Mortality Rate (IMR) + Health Expenditure per Capita (HEC) + Health Expenditure as % of GDP (HGDP) + (MCS*LED) + (MCS*IMR) + (MCS*HEC) + (MCS*HGDP)].

ii) The impact of Internet Usage on Health System = *f*[Internet Users (IU) + Life Expectancy at Birth (LEB) + Infant Mortality Rate (IMR) + Health Expenditure per Capita (HEC) + Health Expenditure as % of GDP (HGDP) + (IU*LED) + (IU*IMR) + (IU*HEC) + (IU*HGDP)]

iii) The impact of Main Telephone Line on Health System = *f*[Main Telephone Line (MTL) + Life Expectancy at Birth (LEB) + Infant Mortality Rate (IMR) + Health Expenditure per Capita (HEC) + Health Expenditure as % of GDP (HGDP) + (MTL*LED) + (MTL*IMR) + (MTL*HEC) + (MTL*HGDP)].

## 4 Methodology

The methodology employed in this study involves a multi-method approach in which three analytical techniques – Data Envelopment Analysis (DEA), Cluster Analysis (CA) and Partial Least Squares (PLS) were adopted. The next section presents these methods in detail.

### 4.1 Data Envelopment Analysis (DEA)

Efficiency measurement has been the focus of organisations as well as economies in improving productivity (Cook and Seiford 2009). Farrell (1957) places emphasis on a need to combine the measurements of the multiple inputs into a satisfactory measure of efficiency. DEA entails a principle of extracting information about a population of observations to evaluate efficiency with reference to an imposed efficient frontier. This process occurs when DEA calculates a discrete piecewise frontier determined by a set of referent (efficient) decision making units (DMUs) which are identified by the ability to utilize the same level of inputs and produce same or higher outputs (Coelli 1996; Cooper, Seiford and Zhu 2011). It involves the use of linear programming to calculate a performance measure (efficiency) for each DMU relative to all the other DMUs with regards to the sole requirement that all observations lie on or below the extreme frontier (Cooper et al. 2011).

DEA was initiated by Charnes et al. (1978), following on the earlier work of Farrell (1957). Charnes et al. (1978) proposed a DEA model which assumed constant returns to scale (CRS) and subsequently Banker, Charnes and Cooper (1984) proposed an alternative model known as the variable returns to scale (VRS). DEA model is flexible by which it can estimate input/output in two common orientations: *input-oriented* *and output-oriented.* An input orientation involves the minimisation of inputs to achieve a given level of output while an output orientation is the maximisation of outputs for a given level of inputs (Cooper et al. 2011).

### 4.2 Cluster Analysis (CA)

CA is an exploratory data analysis technique that organises data into groups or clusters in such a way that objects with more similar multivariate characteristics are placed in one group (cluster) than those in other groups. This means that cluster analysis reduces the number of cases or observations by organising them into smaller cases or clusters based on proximity (Burns and Burns 2008).

There are two types of algorithms used to perform cluster analysis: Hierarchical and non-hierarchical. In hierarchical algorithm, a delete- or add-elements operation is performed which build a tree-like structure (Ketchen and Shook 1996). Hierarchical algorithms include agglomerative and divisive methods. Agglomerative methods focus on adding elements to clusters while divisive methods focus on deleting them from clusters. However, hierarchical algorithms are characterised by poor cluster assignment resulting from single pass through data set (Ketchen and Shook 1996).

On the other hand, non-hierarchical algorithms (also known as K-means or iterative methods) partition a data set into a pre-specified number of clusters (K) to arrive at optimal solution or cluster solution (Ketchen and Shook 1996). K-means clustering has advantages and tends to be favoured over hierarchical algorithms. First, K-means clustering is less affected by outlier elements. By performing several passes through the data set, observations switch cluster membership thereby correcting outlier elements (Hair, Anderson, Tatham and Black 1992). Second, K-means clustering has the capability to optimise solution within cluster homogeneity and between cluster heterogeneity - increasing cluster membership homogeneity and increasing between clusters' heterogeneity (Ketchen and Shook 1996).

### 4.3 Partial Least Squares Based Structural Equation Model

The Partial Least Squares (PLS) based Structural Equation Model (SEM) is referred to as a component (variance) based SEM. It is a technique used to estimate the coefficients of structural equations with the Partial Least Squares method (Geladi and Kowalski 1986). The PLS approach (also known as PLS-Path Modeling) was developed by Wold (1966). The approach to PLS consists of two iterative procedures: The use of least squares estimation for single model and multi-component models (Urbach and Ahlemann 2010). These iterative procedures enable the minimisation of the variance of the dependent variables where the cause and effect directions between the variables are defined (Chin 1998). Consequently, the model quality is ascertained as more indicators are used to explain the latent. The advantages of PLS analysis are that it allows for theory confirmation and development in the initial stages while at a later stage, it facilitates the development of propositions by exploring the relationships between variables (Chin 1998). In this study, we adopted the later. There are many characteristics that make PLS attractive such as: (Chin 1998; Urbach and Ahlemann 2010).

- PLS has distribution free features where there are no assumptions regarding the distributional form of measured variables.
- PLS can generate latent variable approximation for all the data set cases
- PLS does not allow for independent observations or identical distribution of residuals

### 4.4 Data Sources

The data for this study were obtained from several archival sources: the International Telecommunication Union - ITU (for ICT infrastructure data), the World Health Organization - WTO (for health services and individual health data). The data were readily available for 27 African countries. The data from these sources have often been used in ICT research. The ITU is one of the United Nations (UN) groups which have the most reliable data for the ICT sector. WHO is a specialised agency of the UN that is concerned with international public health. It provides data for monitoring the global health situation.

## 4.5 Data Analysis

Data analysis involved three stages: first, data envelopment analysis, second, cluster analysis and, third, partial least squares. There stages are further discussed below.

### 4.5.1 First Stage: Development of Data Envelopment Analysis (DEA)

The first step in this design is to select the DMUs required for the investigation. There are factors to consider when selecting DMUs - Homogeneity and number of DMUs (Tyagi, Yadav, and Singh 2009). DMUs must be homogeneous units performing same tasks and have similar objectives. Based on homogeneity, 27 African countries were selected and data for ten years from 1998 to 2007 were obtained. On the other hand, the number of DMUs is expected to be larger than the number of product of inputs and outputs for effective discrimination between efficient and inefficient DMUs (Avkiran 2001). For ICT infrastructure, four inputs and ten outputs were selected (Table 3); and two inputs and five outputs were selected for health systems (Table 4). All these variables were also used in previous studies (see for example, Bankole, Osei-Bryson and Brown, 2011; Bankole, Osei-Bryson and Brown 2015; Jeremic et al. 2012; Lee et al. 2005; Samoilenko 2008).

| | Input variables | | Output variables |
|---|---|---|---|
| 1 | Annual telecom investment (% of GDP-curr US $) | 1 | Annual outgoing call traffic(minutes) |
| 2 | Line capacity of exchanges | 2 | Annual incoming call traffic(minutes) |
| 3 | International Internet bandwidth (Mbps) | 3 | Main telephone line subscribers(per 100 inhabitants) |
| 4 | Full-time telecoms staff (% of total labour force) | 4 | Internet users (per 100 inhabitants) |
| | | 5 | Mobile cellular subscribers(per 100 inhabitants) |
| | | 6 | Main telephone lines in operations |
| | | 7 | Percentage of households with telephone |
| | | 8 | Percentage of digital main line |
| | | 9 | Percentage of residential main line |
| | | 10 | Percentage population coverage of mobile phone |

*Table 3. Input and output variables of the DEA model for ICT investment*

| | Input variables | | Output variables |
|---|---|---|---|
| 1 | Health expenditure per capita current US $ | 1 | Life expectancy at birth |
| 2 | Health expenditure total % GDP | 2 | Female adult mortality rate |
| | | 3 | Male adult mortality rate |
| | | 4 | Infant mortality rate |
| | | 5 | Under five infant mortality rate |

*Table 4. Input and output variables of the DEA model for Health systems*

Based on the above input and output variables, DEA were performed ten times, one each for each of the ten years for each of the 27 countries. Consequently, we had ten relative efficiency scores for each country in our study. We used multiplier model to calculate the relative efficiency scores for both ICT infrastructure and Health systems based on input orientation. These scores refer to the relative efficiency of transforming ICT infrastructure into improved ICT services. Similarly, scores for health systems refer to the relative efficiency of transforming health systems (inputs) into improved health care services.

### 4.5.2 Second Stage: Cluster Analysis (CA)

Cluster analysis was then performed on ICT infrastructure and health systems. The mean relative efficiency scores were used to group countries. This means that, on the one hand, 27 mean relative efficiency scores of ICT infrastructure were used to group the 27 countries into respective groups/clusters.

On the other hand, 27 mean relative efficiency scores of health systems were used to group the 27 countries into respective clusters.

#### 4.5.3 Third Stage: Partial Least Squares (PLS)

PLS was then performed to determine the impact of ICT infrastructure on health system. ICT Infrastructure was represented by the following variables: Mobile cellular subscribers (per 100 inhabitants), Internet users (per 100 inhabitants) and Main telephone line (per 100 inhabitants). On the other hand, health system was represented by the following variables: Life expectancy at birth, Infant mortality rate, Health expenditure per capita and Health expenditure as % of GDP

## 5  Results

### 5.1  Data Envelopment Analysis (DEA) Results

We used MaxDea Basic version 6.4 software to calculate the relative efficiency scores for the 27 countries. Results indicate that only one country (Cape Verde) scored hundred percent indicating that it efficiently transformed ICT infrastructure into ICT services. However, there are also those countries that scored less than hundred percent, indicating that they did not efficiently transform their ICT infrastructure and health expenditures into meaningful ICT and health services respectively. However, researchers have suggested that relative efficiency scores of hundred percent or less is not a good indicator of determining the nature of relative efficiency (Ali 1994; Samoilenko 2008). Thus, it is difficult to determine which countries efficiently transformed their inputs into outputs. This situation demands a second stage analysis.

The mean relative efficiency scores were then calculated to obtain a single relative efficiency score from those 10 relative efficiency scores for each country. The mean efficiency scores were then used as input for the CA as presented next.

### 5.2  Cluster Analysis (CA) Results

Using SPSS software (version 16), a non-hierarchical, K-means clustering algorithm was used to cluster countries according to their means. We started by experimenting six clusters (K=6) then examining the cluster membership homogeneity and between cluster heterogeneity. The process was repeated for five, four and three clusters. A three-cluster solution provided the most significant solution for both ICT infrastructure and Health system. Analysis of variance (ANOVA) and F values indicate that the three formed clusters differ significantly (see Table 5). Cluster 1 formed from the ICT infrastructure consists of 14 countries (cases) and their relative efficiencies were the highest compared to other two clusters. Some of the countries in this cluster are Cape Verde with relative efficiency score of 1 and Sudan with 0.99. Cluster 2 consists of three countries which mostly scored 0.6. This cluster is characterised by the lowest relative efficiency scores of ICT infrastructure. This means that the countries (e.g. Kenya, Mali) in this cluster poorly transformed ICT infrastructure into ICT services. Cluster 3 consists of ten countries which scored the second highest relative efficiency of ICT infrastructure. Their scores swing at about 0.9.

On the other hand, cluster 1 formed from health system consists of eight countries. These countries seem to have performed best in transforming health system inputs into health outputs (services). They have an average score of 0.85. Cluster 2 consists of seven countries with the third best scores at a tune of an average of 0.56. These countries seem to have poorly transformed health system inputs into health services. Lastly, Cluster 3 consists of twelve countries with the second best relative efficiency scores of health system. On average, they scored 0.71.

A further examination of the clusters and their member countries reveals that relative efficiency scored by countries in converting ICT infrastructure (inputs) into ICT services corresponds to those in health system. This means that, in general, countries that efficiently transformed ICT infrastructure into ICT services are the ones that efficiently transformed health system inputs (health expenditure) into health services, and vice versa. This correspondence in cluster analysis results suggests a causal-effect relationship which is analysed and presented in the PLS results.

| ICT Infrastructure | | | Health Systems | | |
|---|---|---|---|---|---|
| Cluster | Country | Relative Efficiency | Cluster | Country | Relative Efficiency |
| 1 | Mauritius | .9875464 | 1 | Mauritius | .8065914 |
| | Egypt | .9891203 | | Algeria | .9111846 |
| | Cape Verde | 1.0000000 | | Botswana | .9042045 |
| | South Africa | .9673626 | | Kenya | .8140821 |
| | Swaziland | .9811845 | | Sudan | .7979735 |
| | Sudan | .9900979 | | Lesotho | .8337950 |
| | Lesotho | .9875810 | | Eritrea | .8800380 |
| | Nigeria | .9652538 | | Ethiopia | .8533401 |
| | Togo | .9855771 | | Cluster 1 Countries N=8 | |
| | Senegal | .9914607 | 2 | Tunisia | .6058385 |
| | Gambia | .9537186 | | Egypt | .6018855 |
| | Ethiopia | .9402519 | | South Africa | .4280078 |
| | Mozambique | .9863180 | | Ghana | .5762350 |
| | Burkina Faso | .9880057 | | Djibouti | .5454059 |
| | Cluster 1 Countries N=14 | | | Uganda | .5947976 |
| 2 | Tunisia | .6768278 | | Togo | .5888817 |
| | Kenya | .6442279 | | Cluster 2 Countries N=7 | |
| | Mali | .6475766 | 3 | Cape Verde | .6654959 |
| | Cluster 2 Countries N=3 | | | Morocco | .6748285 |
| 3 | Algeria | .9181360 | | Swaziland | .7603352 |
| | Botswana | .8518003 | | Madagascar | .6560990 |
| | Morocco | .8797570 | | Nigeria | .7074549 |
| | Madagascar | .8058630 | | Benin | .7702314 |
| | Ghana | .8863167 | | Cote d'Ivoire | .6813410 |
| | Djibouti | .8931628 | | Senegal | .6978128 |
| | Uganda | .9079927 | | Gambia | .7417388 |
| | Benin | .8254656 | | Mozambique | .7450982 |
| | Cote d'Ivoire | .8331294 | | Burkina Faso | .7043629 |
| | Eritrea | .9074659 | | Mali | .7022937 |
| | Cluster 3 Countries N=10 | | | Cluster 3 Countries N=12 | |
| Total 27 Countries | | | Total 27 Countries | | |
| ANOVA | | | ANOVA | | |
| df | F | Sig. | df | F | Sig. |
| 2 | 182.362 | .000 | 2 | 69.706 | .000 |

*Table 5. Clusters Analysis and Analysis of Variance*

### 5.3 Partial Least Squares (PLS) Results

Using WarpPLS software (version 5), we ran the Partial Least Squares based Structural Equation Modeling (PLS-SEM) to examine the impact of ICT infrastructure on Health system. Table 6 presents the results of the PLS.

Findings on the impact of mobile communication services on health system indicate that there is a significant (p<.001) impact of mobile communication services on health system. While mobile communication services significantly impact on life expectancy at birth, health expenditure per capita and health expenditure (as % GDP) in a positive way, they also significantly impact on infant mortality rate in a negative way. One unit (per 100 habitants) increase in mobile communication services tends to increase life expectancy at birth by about 30 percent and reduce infant mortality rate by almost 50 percent as well. The following propositions are therefore suggested:

***P1***: *The usage of mobile communication services in health care services is positively related to improved life expectancy at birth*

*P2*: *The usage of mobile communication services in health care services is positively related to reduced infant mortality rate*
*P3*: *The usage of mobile communication services in health care services is positively related to increased health expenditure per capita*
*P4*: *The usage of mobile communication services in health care services is positively related to increased health expenditure as percentage of GDP*

The results of the present study corroborate the findings of previous studies. For example, Wu and Raghupathi (2012) found that use of ICT in providing health related information to the public in 200 countries improved life expectancy and reduced mortality rate. Similarly, Raghupathi and Wu (2011) found that ICT is positively related to life expectancy at birth and negatively related to mortality rate of a country's population.

Similarly, the findings of the present study that increased use of mobile communication services leads to increased health expenditure is also supported by previous studies such as that of Raghupathi and Wu (2011). However, these findings may seem as a surprise for many who believe that ICT and in particular technologies when used must reduce costs. The plausible explanation is that when mobile communication services are used to propagate health related information such as of vaccination schedules to the public, there will always be an increased purchase of say, more vaccines to cater for the high demand as a result of publicised vaccination schedules. Due to availability of health related information and services afforded by ICT, health intervention such as increase immunisation, reduce infection, reduce mortality and increase life expectancy will surely met at an extra money that would lead to an increase in health expenditure per capita and in that of as percentage of GDP. However, this has been explained as a rebound effect in additional investment due to effects of ICTs (Raghupathi and Wu 2011). This is why ICT seems to lead to a positive increase in health expenditure, which over the long term may decrease due to preventive measures' pay off (Raghupathi and Wu 2011).

ICT has opened up doors for health information flow and allowed for mobile health arrangements to take place. Information which was once difficult to receive in underserved community in the past, it is now getting easier to receive it due to ICT. Health care information and mobile health facilities are serving the rural communities thereby improving individual's health (Kwankam 2004). Many studies have suggested that poor reach of health care facilities especially in Africa where most rural areas are underserved with health care services; mobile communication services can significantly improve provision of these services (Khatun, Sima and Rokshana 2015; Wu and Raghupathi 2012). The findings suggest that if African countries have to increase life expectancy at birth and reduce infant mortality rate, they need to invest in mobile communication services.

| Variable | Life Expectancy at Birth (LEB) β | Health Expenditure per capita (HEC) β | Health Expenditure % GDP (HGDP) β | Infant Mortality Rate (MI) β |
|---|---|---|---|---|
| Mobile communication services(MCS) | 0.292* | 0.749* | 0.306* | -0.490* |
| Internet Usage (IU) | 0.529* | 0.635* | 0.205* | -0.651* |
| Main Telephone Line (MTL) | 0.669* | 0.749* | -0.171** | -0.813* |

*Note: \*p<0.001, \*\*p=0.002*
*Table 6. Results of Partial Least Square Based Structural Equation Modeling (1998 – 2007)*

Findings on the impact of Internet usage on health system indicate a statistically significant (p <.001) relationship. Internet usage tends to increase life expectancy at birth and significantly reduces infant mortality rate. These results are in support of the previous studies. For example, Morawczynski and Ngwenyama (2007) found that investment in ICT significantly increases life expectancy. Similar findings were reported by other studies such as of Ngwenyama, Andoh-Baidoo, Bollou and Morawczynski (2006) and Bankole (2009).

On the other hand, such investment has also a positive significant impact on health expenditure. This suggests that Internet usage does not reduce health expenditure per capita and as of percentage GDP as well. This may seem contrary to what many believe that Internet usage reduces costs. The reasons provided above also apply here. The present findings lend support to the following propositions:

*P5: Internet usage in health care services is positively related to improved life expectancy at birth*
*P6: Internet usage in health care services is positively related to reduced infant mortality rate*
*P7: Internet usage in health care services is positively related to increased health expenditure per capita*
*P8: Internet usage in health care services is positively related to increased health expenditure as percentage of GDP*

Given the findings, African countries need to increase Internet usage if they need to increase life expectancy at birth and reduce infant mortality rate as well. The results suggest that eHealth has a potential to transform the health sector particularly in Africa where health systems are underdeveloped. The small number of doctors may be optimised to provide health services efficiently and to a large segment of the population. African countries should take advantage of an increasing adoption of mobile services in the region (Mimbi, Bankole and Kyobe 2011).

Similarly, findings on the impact of main telephone line on health system indicate a statistically significant (p<.001) relationship. Investment in main telephone line tends to significantly improve life expectancy at birth and significantly reduce infant mortality rate in Africa. In addition, the findings also indicate that investment in main telephone line infrastructure significantly increases health expenditure per capita (see Table 6). But, it reduces health expenditure as percentage of GDP (significant at p=.002). A unit increase in main telephone line has a potential of reducing health expenditure (as % of GDP) by 17 percent. This is a substantial savings that can be used for other developmental projects. However, these findings need careful interpretation. Given the trend of low availability of main telephone lines in most rural areas in Africa, it might be that few people had access to these telecommunication facilities thereby limiting their uptake of health care services. This seems to be a plausible reason as to why the reduction of 17 percent in health expenditure is recorded as demand for health care services subsided. The following propositions are therefore suggested:

*P9: Usage of main telephone line in health care services is positively related to improved life expectancy at birth*
*P10: Usage of main telephone line in health care services is positively related to reduced infant mortality rate*
*P11: Usage of main telephone line in health care services is positively related to increased health expenditure per capita*
*P12: Usage of main telephone line in health care services is positively related to reduced health expenditure as percentage of GDP*

The overall results suggest that investment in ICT infrastructure improves health of individuals in Africa. However, such investment has a significant impact on health expenditure in Africa at least in short term. Only Internet usage, among the three, has a least impact on health expenditure (as % GDP). On the other hand, main telephone line seems to have a highest impact on health system components examined. Many other studies have also indicated that ICT improves health care services delivery (Khatun et al. 2015; Lucas 2008). The findings of the present study suggest that ICT has the potential to improve people's health through availability of information afforded by ICT. It is estimated that 80 per cent of all children who have died in Africa would have been saved because the knowledge to save them existed but simply could not be availed when and where it was needed (Kwankam 2004).

## 6   Contribution of the Study

The present study has several implications: First, given the existing handful scholarly literature regarding ICT impact on health systems, this study contributes to the body of knowledge in this regard. This study has integrated ICT and public health thereby contributing to interdisciplinary approach to ICT for development research area.  Second, the multi-method approach adopted in the present study to demonstrate the impact of ICT on health systems forms a methodological contribution. The study adopted and implemented DEA at the first stage, CA at the second stage and PLS at the third stage. Each of these analyses aimed at unraveling relationship and data attributes at different stages of analysis.  Third, the present results have implications for policy makers. The findings indicate that efficiency in transforming ICT infrastructure into ICT services has a direct relationship with that of transforming health system inputs into health care outcomes. This means that improved ICT services as a result of well transformed ICT inputs would lead to improved health outcomes. The study has also demonstrated that countries that efficiently transformed ICT inputs also efficiently transformed health system inputs into better ICT and better health care services respectively. The opposite is also true. This calls for attention of policy makers to efficiently transform not only ICT inputs but also health system inputs if ICT is to be leveraged in health systems. The findings of the present study also suggest that investment in ICT is tax-payers money well spent.

## 7   Conclusion

The present study was set to investigate the impact of ICT on health systems in Africa. In the recent past, development actors such as WHO indicated the existence of a link between ICT and health systems. Using a multi-method approach, this study investigates the impact of ICT on health systems in 27 African countries for the period from 1998 to 2007. We were able to demonstrate that ICT has a significant impact on health systems. Specifically, by investing in ICT infrastructure, countries can significantly increase life expectancy at birth and reduce infant mortality rate. However, among the investigated ICT components, main telephone line seems to provide the highest return (impact) by not only increasing the highest level of life expectancy at birth and reducing the highest level of infant mortality rate, but also by reducing health expenditure (as % GDP). It is also important to note that while ICT tends to improve health care services, investment in health systems is also important to compliment ICT investment. It is natural that when ICT is used in health improvement initiatives, there will be, on the short term, increased health expenditure due to the rising health awareness and subsequent high demand of health care services (afforded by ICT) of communities. The lesson learned from the Ebola outbreak in West Africa is a demonstration for the need to improve health systems in Africa. Without health people, social and economic development would not be realised. The nation of ill people will not be able to implement poverty reduction strategies as more useful resources would be used to treat illness.

Besides its contribution, our study has limitations. First, while the data used for the present study provides plausible results, the fast development of ICT and its adoption in Africa may have substantially progressed after 2007. This means that the magnitude of impact of ICT on health systems might have changed as well. Second, the relative efficiency used as clustering criteria produced a three cluster solution. However, if different input variables are used as clustering criteria, the subsequent cluster solution might differ with the one in the present study.

Future research in this area can focus on addressing the above limitations and possibly incorporate more countries and more recently available data. It is also recommended that investigation into reasons as to why other countries efficiently transformed ICT and health system inputs while others did not be conducted. Such studies would have practical implications for finding better ways of utilisation of scarce resources efficiently for improved people's welfare. Finally, the propositions developed in the present study provide an avenue for further investigation.

# 8　References